\documentstyle[prl,aps,preprint]{revtex}
\begin{document}
\bigskip\ 

\begin{center}
{\bf \-MATROID THEORY AND CHERN-SIMONS}

\bigskip\ 

\smallskip\ 

J. A. Nieto\footnote[1]{%
nieto@uas.uasnet.mx}and M. C. Mar\'{\i}n

\smallskip\ 

{\it Facultad de Ciencias F\'{\i}sico-Matem\'{a}ticas de la Universidad
Aut\'{o}noma}

{\it de Sinaloa, 80010 Culiac\'{a}n Sinaloa, M\'{e}xico}

\bigskip\ 

\bigskip\ 

{\bf Abstract}
\end{center}

It is shown that matroid theory may provide a natural mathematical framework
for a duality symmetries not only for quantum Yang-Mills physics, but also
for M-theory. Our discussion is focused in an action consisting purely of
the Chern-Simons term, but in principle the main ideas can be applied beyond
such an action. In our treatment the theorem due to Thistlethwaite, which
gives a relationship between the Tutte polynomial for graphs and Jones
polynomial for alternating knots and links, plays a central role. Before
addressing this question we briefly mention some important aspects of
matroid theory and we point out a connection between the Fano matroid and
D=11 supergravity. Our approach also seems to be related to loop solutions
of quantum gravity based in Ashtekar formalism.

\bigskip\ 

Pacs numbers: 04.60.-m, 04.65.+e, 11.15.-q, 11.30.Ly

May, 2000

\newpage\ 

\noindent {\bf 1.- INTRODUCTION}

\bigskip 

In the last few years, duality has been a source of great interest to study
nonperturbative, as well as perturbative, dynamics of superstrings [1] and
supersymmetric Yang-Mills [2]. In fact, duality is the key physical concept
that relates the five known superstring theories in 9+1 dimensions (i.e.
nine space and one time), Type I, Type IIA, Type IIB, Heterotic SO(32) and
Heterotic E$_8\times $ E$_8$, which may now be understood as different
manifestations of one underlying unique theory called M-theory [3]-[9].
However, dualities are still a mystery and up to now a general understanding
how these dualities arises is missing. Nevertheless, just as the equivalence
principle is a basic principle in general relativity, the recent importance
of dualities in gauge field theories and string theories strongly suggest a
duality principle as a basic principle in M-theory. In this sense, what it
is needed is a mathematical framework to support such a duality principle.

M-theory is defined as a 10+1 dimensional theory arising as the
strong-coupling limit of type IIA string theory. Essentially, M-theory is a
non-pertubative theory and in addition to the five superstring theories it
describes supermembranes [10], 5-branes [11] and D=11 supergravity [12].
Although the complete M-theory is unknown there are two main proposed routes
to construct it. One is the N=(2,1) superstring theory [13] and the other
M(atrix)-theory [14]. Martinec [15] has suggested that these two scenarios
may, in fact, be closely related. This scenario has been extended [16] to
include dualities involving compactifications on timelike circles as well as
spacelike circles ones. In particular, it has been shown that T-duality on a
timelike circle takes type IIA theory into a type IIB$^{*}$ theory and type
IIB$^{*}$ theory into a type IIA theory and that the strong-coupling limit
of type IIA$^{*\text{ }}$is a theory in 9+2 dimensional theory, denoted by M$%
^{*}$.

More recently, Khoury and Verlinde [17] have shed some new light on the old
idea of open/closed string duality [18]. This duality is of special interest
because emphasizes the idea that closed string dynamics (gravity) is dual to
open string dynamics (gauge theory). Two previous examples on this direction
are matrix theory [14], where gravity arises as an effect of open string
quantum fluctuations and Maldacena's conjecture [19] that anti-deSitter
supergravity is in some sense dual to supersymmetric gauge theory.

Thus, just as the tensor theory makes mathematical sense of the postulate of
relativity ``the laws of physics are the same for every observer'', we are
pursuing the possibility that the mathematical formalism necessary to make
sense of a duality principle in M-theory is matroid theory [20]. This theory
is a generalization of matrices and graphs and , in contrast to graphs in
which duality can be defined only for planar graphs, it has the remarkable
property that duality can be defined for every matroid. Since
M(atrix)-theory and N=(2,1) superstrings have had an important success on
describing some essential features of M-theory a natural question is to see
whether matroid theory is related to these two approaches. As a first step
in this direction we may attempt to see if matroid theory is linked somehow
to D=11 supergravity which is a common feature of both formalism. In fact,
it has been shown [21] that the Fano matroid and its dual are closely
related to Englert's compactification [22] of D=11 supergravity. This result
is physically interesting because such a relation allows the connection
between the fundamental Fano matroid or its dual [23] and octonions which,
at the same time, are one of the alternative division algebras [24]. It is
worth mentioning that some time ago Blencowe and Duff [25] raised the
question whether the four forces of nature correspond to the four divisions
algebras.

In this work, we make further progress on this program. Specifically, we
find a route to incorporate matroid theory in quantum Yang-Mills in the
context of Chern-Simons action. Our mechanism is based on a theorem due to
Thistlethwaite [26] which connect the Jones polynomial for alternating knots
with the Tutte polynomial for graphs. Since Witten [27] showed that Jones
polynomial can be understood in three dimensional terms through a
Chern-Simons formalism it became evident that we have a bridge between
graphs and Chern-Simons. In this context duality, which is the main subject
in graphic matroids, can be associated to Chern-Simons in a mathematical
natural way. This connection may transfer important theorems from matroid
theory to fundamental physics. For instance, the theorem due to Whitney [20]
that if $M_1,..,M_p$ and $M_1^{^{\prime }},..,M_p^{^{\prime }}$ are the
components of the matroids $M$ and $M^{\prime }$ respectively, and if $%
M_i^{^{\prime }}$ is the dual of $M_i$ ($i=1,...,p$) then $M^{\prime }$ is
dual of $M$ and conversely, if $M$ and $M^{\prime }$ are dual matroids then $%
M_i^{^{\prime }}$ is dual of $M_i$ may be applied to dual Chern-Simons
partition functions. One of the aims of this work is to explain how this can
be done.

The plan of this work is as follows. In section 2, we briefly review matroid
theory and in section 3 we closely follow the reference [21] to discuss a
connection between matroid theory and D=11 supergravity. In section 4, we
study the relation between matroid theory and Witten's partition function
for knots. Finally, in section 5, we make some final comments.

\smallskip\ 

\noindent {\bf 2.- A BRIEF REVIEW OF MATROID THEORY}

\bigskip 

In 1935, while working on abstract properties of linear dependence, Whitney
[20] introduced the concept of matroid. In the same year, Birkhoff [28]
established the connection between simple matroids (also known as
combinatorial geometries [29]) and geometric lattices. In 1936, Mac Lane
[30] gave an interpretation of matroids in terms of proyective geometry. And
an important progress to the subject was given in 1958 by Tutte [23] who
introduced the concept of homotopy for matroids. At present, there is a
large body of information about matroid theory. The reader interested in the
subject may consult the excellent books on matroid theory by Welsh [31],
Lawler [32] and Tutte [33]. We also recommend the books of Wilson [34], Kung
[35] and Ribnikov [36].

An interesting feature of matroid theory is that there are many different
but equivalent ways of defining a matroid. In this respect, it seems
appropriate to briefly review the Whitney's [20] discovery of the matroid
concept. While working with linear graphs Whitney noticed that for certain
matrices duality had a simple geometrical interpretation quite different
than in the case of graphs. Further, he observed that any subset of columns
of a matrix is either linearly independent or linearly dependent and that
the following two theorems must hold:

(a) any subset of an independent set is independent.

(b) if N$_p$ and N$_{p+1}$ are independent sets of p and p+1 columns
respectively, then N$_p$ together with some column of N$_{p+1}$ forms an
independent set of p+1 columns.

\noindent Moreover, he discovered that if these two statements are taking as
axioms then there are examples that do not represent any matrix and graph.
Thus, he concluded that a system satisfying (a) and (b) should be a new one
and therefore deserved a new name: He called to this kind of system a
matroid.

The definition of a matroid in terms of independent sets has been refined
and is now expressed as follows: A matroid $M$ is a pair (E,${\cal I)}$,
where E is a non-empty finite set, and ${\cal I}$ is a non-empty collection
of subsets of E (called independent sets) satisfying the following
properties:

(${\cal I}$ {\it i) }any subset of an independent set is independent;

(${\cal I}${\it \ ii) }if I and J are independent sets with I$\subseteq $ J,
then there is an element $e$ contained in J but not in I, such that I$\cup
\{e\}$ is independent.

A base is defined to be any maximal independent set. By repeatedly using the
property (${\cal I}$ {\it ii) }it is straightforward to show that any two
bases have the same number of elements. A subset of E is said to be
dependent if it is not independent. A minimal dependent set is called a
circuit. Contrary to the bases not all circuits of a matroid have the same
number of elements.

An alternative definition of a matroid in terms of bases is as follows:

A matroid $M$ is a pair (E, ${\cal B}$), where E is a non-empty finite set
and ${\cal B}$ is a non-empty collection of subsets of E (called bases)
satisfying the following properties:

(${\cal B}$ {\it i) }no base properly contains another base;

(${\cal B}$ {\it ii)} if B$_1$ and B$_2$ are bases and if $b$ is any element
of B$_1,$ then there is and element $g$ of B$_2$ with the property that (B$%
_1 $-\{$b$\})$\cup \{g\}$ is also a base.

A matroid can also be defined in terms of circuits:

A matroid $M$ is a pair (E, ${\cal C})$, where E is a non-empty finite set,
and ${\cal C}$ is a collection of a non-empty subsets of E (called circuits)
satisfying the following properties.

(${\cal C}$ {\it i}) no circuit properly contains another circuit;

(${\cal C}$ {\it ii) }if ${\cal C}_1$ and ${\cal C}_2$ are two distinct
circuits each containing and element $c$, then there exists a circuit in $%
{\cal C}_1$ $\cup $ ${\cal C}_2$ which does not contain $c$.

If we start with any of the three definitions the other two follows as a
theorems. For example, it is possible to prove that (${\cal I}${\it \ )}
implies (${\cal B}$) and (${\cal C}$). In other words, these three
definitions are equivalent. There are other definitions also equivalent to
these three, but for the purpose of this work it is not necessary to
consider them.

Notice that even from the initial structure of a matroid theory we find
relations such as independent-dependent and base-circuit which suggests
duality. The dual of $M$, denoted by $M^{*},$ is defined as a pair (E, $%
{\cal B}^{*}$), where ${\cal B}^{*}$ is a non-empty collection of subsets of
E formed with the complements of the bases of M. An immediate consequence of
this definition is that every matroid has a dual and this dual is unique. It
also follows that the double-dual $M^{**}$ is equal to $M$. Moreover, if A
is a subset of E, then the size of the largest independent set contained in
A is called the rank of A and is denoted by $\rho $(A). If $M=M_1+M_2$ and $%
\rho $($M$) = $\rho $($M_1$) +$\rho $($M_2$) we shall say that $M$ is
separable. Any maximal non-separable part of $M$ is a component of $M$. An
important theorem due to Whitney [20] is that if $M_1,..,M_p$ and $%
M_1^{^{\prime }},..,M_p^{^{\prime }}$ are the components of the matroids $M$
and $M^{\prime }$ respectively, and if $M_i^{^{\prime }}$ is the dual of $%
M_i $ (i = 1,...,p). Then $M^{\prime }$ is dual of $M$. Conversely, let $M$
and $M^{\prime }$ be dual matroids, and let $M_1,..,M_p$ be components of $M$%
. Let $M_1^{^{\prime }},..,M_p^{^{\prime }}$ be the corresponding
submatroids of $M^{\prime }$. Then $M_1^{^{\prime }},..,M_p^{^{\prime }}$
are the components of $M^{\prime }$, and $M_i^{^{\prime }}$ is dual of $M_i.$

\smallskip\ 

\noindent {\bf 3.- MATROID THEORY AND SUPERGRAVITY}

\bigskip 

Among the most important matroids we find the binary and regular matroids. A
matroid is binary if it is representable over the integers modulo two. Let
us clarify this definition. An important problem in matroid theory is to see
which matroids can be mapped in some set of vectors in a vector space over a
given field. When such a map exists we are speaking of a coordinatization
(or representation) of the matroid over the field. Let GF(q) denote a finite
field of order q. Thus, we can express the definition of a binary matroid as
follows: A matroid which has a coordinatization over GF(2) is called binary.
Furthermore, a matroid which has a coordinatization over every field is
called regular. It turns out that regular matroids play a fundamental role
in matroid theory, among other things, because they play a similar role that
planar graphs in graph theory [34]. It is known that a graph is planar if
and only if it contains no subgraph homeomorphic to K$_5$ or K$_{3,3}$. The
analogue of this theorem for matroids was provided by Tutte [23]. In fact,
Tutte showed that a matroid is regular if and only if is binary and includes
no Fano matroid or the dual of this. In order to understand this theorem it
is necessary to define the Fano matroid. We shall show that the Fano matroid
may be connected with octonions which, in turn, are related to the Englert's
compactification of D=11 supergravity.

A Fano matroid F is the matroid defined on the set E=\{1,2,3,4,5.6.7\} whose
bases are all those subsets of E with three elements except f$_1=$\{1,2,4\},
f$_2=$\{2,3,5\}, f$_3=$ \{3,4,6\}, f$_4=$\{4,5,7\}, f$_5=$\{5,6,1\}, f$_6=$%
\{6,7,2\} and f$_7=$\{7,1,3\}. The circuits of the Fano matroid are
precisely these subsets and its complements. It follows that these circuits
define the dual F$^{*}$ of the Fano matroid.

Let us write the set E in the form{\it \ }${\cal E}$=$%
\{e_1,e_2,e_{3,}e_{4,}e_5,e_6,e_7\}$. Thus, the subsets used to define the
Fano matroid now become ${\it f}_1=\{e_1,e_2,e_4\}$, ${\it f}%
_2\{e_2,e_3,e_5\}$, ${\it f}_3\{e_3,e_4,e_6\}$, ${\it f}_4\{e_4,e_5,e_7\}$, $%
{\it f}_5\{e_5,e_6,e_1\}$, ${\it f}_6\{e_6,e_7,e_2\}$ and ${\it f}_7$ $%
\{e_7,e_1,e_3\}$. The central idea is to identify the quantities $e_i,$
where $i=1,...$,$7,$ with the octonionic imaginary units. Specifically, we
write an octonion $q$ in the form $%
q=q_0e_0+q_1e_1+q_2e_2+q_3e_3+q_4e_4+q_5e_5+q_6e_6+q_7e_7,$ where $q_0$ and $%
q_i$ are real numbers. Here, $e_0$ denotes the identity. The product of two
octonions can be obtained with the rule:

\begin{equation}
e_ie_j=-\delta _{ij}+\psi _{ij}^ke_k,  \label{ec.1}
\end{equation}
where $\delta _{ij}$ is the Kronecker delta and $\psi _{ijk}=$ $\psi
_{ij}^l\delta _{lk}$ is the fully antisymmetric structure constants, with $%
i,j,k=1,...,7$. By taking the $\psi _{ijk}$ equals 1 for one of the seven
combinations ${\it f}_i$ we may derive all the values of $\psi _{ijk}$.

The octonion (Cayley) algebra is not associative, but alternative. This
means that the basic associator of any three imaginary units is

\begin{equation}
(e_i,e_j,e_k)=(e_ie_j)e_k-e_i(e_je_k)=\varphi _{ijkm}e_m,  \label{ec.2}
\end{equation}
where $\varphi _{ijkl}$ is a fully antisymmetric four index tensor. It turns
out that $\varphi _{ijkl}$ and $\psi _{ijk}$ are related by the expression

\begin{equation}
\varphi _{ijkl}=(1/3!)\epsilon _{ijklmnr}\psi _{mnr},  \label{ec.3}
\end{equation}
where $\epsilon _{ijklmnr}$ is the completely antisymmetric Levi-Civita
tensor, with $\epsilon _{12...7}=1$. It is interesting to note that given
the numerical values f$_i$ for the indices of $\psi _{mnr}$ and using (3) we
get the other seven subsets of E with four elements of the dual Fano matroid
F$^{*}.$ For instance, if we take f$_1$ then we have $\psi _{124}$ and (3)
gives $\varphi _{3567}$ which leads to the circuit subset $\{3,5,6,7\}$.

We would like now to relate the above structure to the Englert's octonionic
solution [22] of eleven dimensional supergravity. First, let us introduce
the metric

\begin{equation}
g_{ab}=\delta _{ij}h_a^ih_b^j,  \label{ec.4}
\end{equation}
where $h_a^i$ = $h_a^i(x^c)$ is a sieben-bein, with $a,b,c=1,...,7$. Here, $%
x^c$ are a coordinates patch of the geometrical seven sphere S$^7$. The
quantities $\psi _{ijk}$ can now be related to the S$^7$ torsion in the form

\begin{equation}
T_{abc}=R_0^{-1}\psi _{ijk}h_a^ih_b^jh_c^k,  \label{ec.5}
\end{equation}
where $R_0$ is the S$^7$ radius. While the quantities $\varphi _{ijkl}$ can
be identified with the four index gauge field $F_{abcd}$ through the formula

\begin{equation}
F_{abcd}=R_0^{-1}\varphi _{ijkl}h_a^ih_b^jh_c^kh_d^l.  \label{ec.6}
\end{equation}
Furthermore, it is possible to prove that the Englert's 7-dimensional
covariant equations are solve with the identification $F_{abcd}=\lambda
T_{[abc},_{d]},$ where $\lambda $ is a constant. Therefore, $\lambda
T_{abc}=A_{abc}$ is the fully antisymmetric gauge field which is a
fundamental object in 2-brane theory [6].

It is important to mention that in the Englert's solution of D= 11
supergravity the torsion satisfies the Cartan-Schouten equations

\begin{equation}
T_{acd}T_{bcd}=6R_0^{-2}g_{ab},  \label{ec.7}
\end{equation}

\begin{equation}
T_{ead}T_{dbf}T_{fce}=3R_0^{-2}T_{abc.}  \label{ec.8}
\end{equation}
But as Gursey and Tze [37] noted, these equations are mere septad-dressed,
i.e. covariant forms of the algebraic identities

\begin{equation}
\psi _{ikl}\psi _{jkl}=6\delta _{ij},  \label{ec.9}
\end{equation}

\begin{equation}
\psi _{lim}\psi _{mjn}\psi _{nkl}=3\psi _{ijk},  \label{ec.10}
\end{equation}
respectively. It is worth mentioning that Englert solution realizes the
riemannian curvature-less but torsion-full Cartan-geometries of absolute
parallelism on S$^7$.

So, we have shown that the Fano matroid is closely related to octonions
which at the same time are an essential part of the Englert's solution of
absolute parallelism on S$^7$ of D=11 supergravity. The Fano matroid and its
dual are the only minimal binary irregular matroids. We know from Hurwitz
theorem (see reference [24]) that octonions are one of the alternative
division algebras (the others are the real numbers, the complex numbers and
the quaternions). While among the only parallelizable spheres we find S$^7$
(the other are the spheres S$^1$ and S$^3$ [38]). This distinctive and
fundamental role played by the Fano matroid, octonions and S$^7$ in such
different areas in mathematics as combinatorial geometry, algebra and
topology respectively lead us to believe that the relationship between these
three concepts must have a deep significance not only in mathematics, but
also in physics. Of course, it is known that the parallelizability of S$^1$,
S$^3$ and S$^7$ has to do with the existence of the complex numbers, the
quaternions and the octonions respectively (see reference [39]). It is also
known that using an algebraic topology called K-theory [40] we find that the
only dimensions for division algebras structures on Euclidean spaces are 1,
2, 4, and 8. We can add to these remarkable results another fundamental
concept in matroid theory; the Fano matroid.

\smallskip\ 

\noindent {\bf 4.- MATROID THEORY AND CHERN-SIMONS}

\bigskip 

Before going into details, it turns out to be convenient to slightly modify
the notation of the previous section. In this section, we shall assume that
the Greek indices $\alpha ,\beta ,...,etc$ run from 0 to 3, the indices $%
i,j,...,etc$ run from 0 to 2 and finally the indices $a,b,...,etc$ take
values in the rank of a compact Lie Group G. Further, we shall denote a
compact oriented four manifold as $M^4.$

Consider the second Chern class action

\begin{equation}
S=\frac k{16\pi }\int_{M^4}\epsilon ^{\mu \nu \alpha \beta }F_{\mu \nu
}^aF_{\alpha \beta }^b\text{ }g_{ab},  \label{11}
\end{equation}
with the curvature given by

\begin{equation}
F_{\alpha \beta }^a=\partial _\alpha A_\beta ^a-\partial _\beta A_\alpha
^a+C_{bc}^aA_\alpha ^bA_\beta ^c.  \label{12}
\end{equation}
Here $g_{ab}$ is the Killling-Cartan metric and $C_{bc}^a$ are the
completely antisymmetric structure constants associated to the compact
simple Lie group G. The action (11) is a total derivative and leads to the
Chern-Simons action

\begin{equation}
S_{CS}=\frac k{4\pi }\int_{M^3}\{\epsilon ^{ijk}(A_i^a(\partial
_jA_k^b-\partial _kA_j^b)g_{ab}+\frac 23C_{abc}A_i^aA_j^bA_k^c)\},
\label{13}
\end{equation}
where $M^3=\partial M^4$ is a compact oriented three dimensional manifold.
In a differential forms notation (13) can be rewritten as follows:

\begin{equation}
S_{CS}=\frac k{2\pi }\int_{M^3}Tr(A\wedge dA+\frac 23A\wedge A\wedge A),
\label{14}
\end{equation}
where $A=A_i^aT_adx^i$, with $T_a$ the generators of the Lie algebra of G.

Given a link $L$ with $r$ components and irreducible representation $\rho _r$
of G, one for each component of the link, Witten [27] defines the partition
function

\begin{equation}
Z(L,k)=\smallint D{\cal A}\text{exp(}iS_{cs})\prod\limits_{r=1}^nW(L_r\text{,%
}\rho _r),  \label{15}
\end{equation}
where $W(C_i,\rho _i)$ is the Wilson line

\begin{equation}
W(L_r,\rho _r)=Tr_{\rho _r}P\exp (\smallint _{L_r}A_i^aT_adx^i).  \label{16}
\end{equation}
Here the symbol $P$ means the path-ordering along the knots $L_r.$

If we choose $M^3=S^3,$ $G=SU(2)$ and $\rho _r$=C$^2$ for all link
components then the Witten's partition function (15) reproduces the Jones
polynomial

\begin{equation}
Z(L,k)=V_L(t),  \label{17}
\end{equation}
where

\begin{equation}
t=e^{\frac{2\pi i}k}.  \label{18}
\end{equation}
Here $V_L(t)$ denotes the Jones polynomial satisfying the skein relation:

\begin{equation}
t^{-1}V_{L_{+}}-tV_{L_{-}}=(\sqrt{t}-\frac 1{\sqrt{t}})V_{L_0},  \label{19}
\end{equation}
where $L_{+},L_{-}$ and $L_0$ are the standard notation for overcrossing,
undercrossing and zero crossing.

Now, lets pause about the relation between the knots and Chern-Simons term
and let us discuss the Tutte polynomial. To each graph ${\cal G}$, we
associate a polynomial $T_{{\cal G}}(x,x^{-1})$ with the property that if $%
{\cal G}$ is composed solely of isthmus and loops then $T_{{\cal G}%
}(x,x^{-1})=x^Ix^{-l},$ where $I$ is the number of isthmuses and $l$ is the
number of loops. The polynomial $T_{{\cal G}}$ satisfies the skein relation

\begin{equation}
T_{{\cal G}}=T_{{\cal G}^{^{\prime }}}+T_{{\cal G}^{^{\prime \prime }}},
\label{20}
\end{equation}
where ${\cal G}^{\prime }$ and ${\cal G}^{\prime \prime }$ are obtained by
delating and contracting respectively an edge that is neither a loop nor an
isthmus of ${\cal G}$.

There is a theorem due to Thistlethwaite [26] which assures that if $L$ is
an alternating link and ${\cal G}(L)$ the corresponding planar graph then
the Jones polynomial $V_L(t)$ is equal to the Tutte polynomial $T_{{\cal G}%
}(-t,-t^{-1})$ up to a sign and factor power of $t.$ Specifically, we have

\begin{equation}
V_L(t)=(-t^{\frac 34})^{w(L)}t^{\frac{-(r-n)}4}T_{{\cal G}}(-t,-t^{-1})
\label{21}
\end{equation}
where $w(L)$ is the writhe and $r$ and $n$ are the rank and the nullity of $%
{\cal G}$ respectively$.$ Here $V_L(t)$ is the Jones polynomial of
alternating link $L.$

On the other hand, a theorem due to Tutte allows to compute $T_{{\cal G}%
}(-t,-t^{-1})$ from the maximal trees of ${\cal G}$. In fact, Tutte proved
that if ${\cal B}$ denotes the maximal trees in a graph ${\cal G}$, $i(B)$
denotes the number of internally active edges in ${\cal G}$ and $e(B)$ the
number the externally active edges in ${\cal G}$ (with respect to a given
maximal tree $B\epsilon {\cal B}$) then the Tutte polynomial is given by the
formula

\begin{equation}
T_{{\cal G}}(-t,-t^{-1})=\sum x^{i(B)}x^{-e(B)}  \label{22}
\end{equation}
where the sum is over all elements of ${\cal B}$.

First, note that ${\cal B}$ is the collection of bases of ${\cal G}$. If we
now remember our definition of matroid $M$ in terms of bases discussed in
section 2 we note the Tutte polynomial $T_{{\cal G}}(-t,-t^{-1})$ computed
according to (22) uses the concept of a graphic matroid $M({\cal G)}$
defined as the pair (E, ${\cal B}$), where E is the set of edges of ${\cal G}
$. In fact, the elements of ${\cal B}$ satisfy the two properties

(${\cal B}$ {\it i) }no base properly contains another base;

(${\cal B}$ {\it ii)} if $B_1$ and $B_2$ are bases and if $b$ is any element
of $B_1,$ then there is and element $g$ of $B_2$ with the property that ($%
B_1 $-\{$b$\})$\cup \{g\}$ is also a base.

\noindent which identifies a $M({\cal G)}$ as a matroid. With this
remarkable connection between the Tutte polynomial and a matroid we have
found in fact a connection between the partition function $Z(L,k)$ given in
(15) and matroid theory. This is because according to (21) the Tutte
polynomial $T_{{\cal G}}(-t,-t^{-1})$ are related to the Jones polynomial $%
V_L(t)$ which at the same time according to (17) are related to the
partition function $Z(L,k)$. Specifically, for $M^3=S^3,$ $G=SU(2)$, $\rho _r
$=C$^2$ for all alternating link components of $L$, we have the relation

\begin{equation}
Z(L,k)=V_L(t)=(-t^{\frac 34})^{w(L)}t^{\frac{-(r-n)}4}T_{{\cal G}%
}(-t,-t^{-1}).  \label{23}
\end{equation}
Thus, the matroid (E, ${\cal B}$) used to compute $T_{{\cal G}}(-t,-t^{-1})$
can be associated not only to $V_L(t),$ but also to $Z(L,k).$

Now that we have at hand this slightly but important connection between
matroid theory and Chern-Simons theory we arre able to transfer information
from matroid theory to Chern-Simons and conversely from Chern-Simons to
matroid theory. Let us discuss two examples for the former possibility.

First of all, it is known that in matroid theory the concept of duality is
of fundamental importance. For example, there is a remarkable theorem that
assures that every matroid has a dual. So, the question arises about what
are the implications of this theorem in Chern-Simons formalism. In order to
address this question let us first make a change of notation $T_{{\cal G}%
}(-t,-t^{-1})\rightarrow T_{M({\cal G)}}(t)$ and $Z(L,k)\rightarrow Z_{M(%
{\cal G)}}(k).$ The idea of this notation is to emphasize the connection
between matroid theory, Tutte polynomial and Chern- Simons partition
function. Consider the planar dual graph ${\cal G}^{*}$ of ${\cal G}$. In
matroid theory we have $M({\cal G}^{*}{\cal )}$ =$M^{*}({\cal G)}$.
Therefore, the duality property of the Tutte polynomial

\begin{equation}
T_{{\cal G}}(-t,-t^{-1})=T_{{\cal G}^{*}}(-t^{-1},-t)  \label{24}
\end{equation}
can be expressed as

\begin{equation}
T_{M({\cal G)}}(t)=T_{M^{*}({\cal G)}}(t^{-1})  \label{25}
\end{equation}
and consequently from (23) we discover that for the partition function $Z_{M(%
{\cal G)}}(k)$ we should have the duality property

\begin{equation}
Z_{M({\cal G)}}(k)=Z_{M^{*}({\cal G)}}(-k).  \label{26}
\end{equation}
This duality symmetry for the partition function $Z_{M({\cal G)}}(k)$ is not
really new, but is already known in the literature as mirror image symmetry
(see, for instance [41], and references quoted there). However, what seems
to be new is the way we had derived it.

As a second example let us first mention another theorem due to Withney
[20]: If $M_1,..,M_p$ and $M_1^{^{\prime }},..,M_p^{^{\prime }}$ are the
components of the matroids $M$ and $M^{\prime }$ respectively, and if $%
M_i^{^{\prime }}$ is the dual of $M_i$ $(i=1,...,p)$. Then $M^{\prime }$ is
dual of $M$. Conversely, let $M$ and $M^{\prime }$ be dual matroids, and let 
$M_1,..,M_p$ be components of $M$. Let $M_1^{^{\prime }},..,M_p^{^{\prime }}$
be the corresponding submatroids of $M^{\prime }$. Then $M_1^{^{\prime
}},..,M_p^{^{\prime }}$ are the components of $M^{\prime }$, and $%
M_i^{^{\prime }}$ is dual of $M_i.$ Thus, according to (26) we find that

\begin{equation}
Z_{M_i({\cal G}_i{\cal )}}(k)=Z_{M_i^{^{\prime }}({\cal G}_i{\cal )}}(-k)
\label{27}
\end{equation}
if and only if

\begin{equation}
Z_{M({\cal G)}}(k)=Z_{M^{\prime }({\cal G)}}(-k),  \label{28}
\end{equation}
where ${\cal G}_i$ are the components of ${\cal G}.$

\smallskip\  

\noindent {\bf 5.- COMMENTS}

\bigskip 

Motivated by a possible duality principle in M-theory we have started to
bring information from matroid theory to fundamental physics. We now have
two good examples which indicate that this task makes sense. In the first
example, we have found enough evidence for a connection between the Fano
matroid and supergravity in D=11. While in the second example, we have found
a relation between the graphic matroid and the Witten's partition function
for Chern-Simons. This relation is of special importance because leads us to
a duality symmetry in the partition function $Z_{M({\cal G)}}(k)$. In fact,
if there is a duality principle in M-theory we should expect to a have a
duality symmetry in the corresponding partition function associated to
M-theory.

In this work, we have concentrated in the original connection between
Chern-Simons action and knots theory. But it is well known that Chern-Simons
formalism and knots connection has a number of extensions [41]. It will be
interesting to study such extensions from the point of view of matroid
theory. It is also known Chern-Simons formalism is closely related to
conformal field theory and this in turn is closely related to string theory.
So, it seems that the present work may eventually leads to a connection
between matroid theory and string theory. In order to achieve this goal we
need to study the relation between matroids and Chern-Simons using signed
graphs [42]. This is because general knots and links (not only alternating)
are one to one correspondence with signed planar graphs. This in turn are
straightforward connected with Kauffmann polynomials [43] which at the same
time leads to the Jones polynomials. But, signed graphs leads to signed
matroids. So, one of our future goals will be to find a connection between
signed matroids and string theory. Moreover, matrix Chern-Simons theory [44]
has a straightforward relation with Matrix-model and non-commutative
geometry [45]. So, a natural extension of the present work will be to study
the relation between matroid theory and matrix Chern-Simons theory.

An important duality in M-theory is the strong/weak coupling $S$-duality
[46] which provides with one of the most important techniques to study
non-perturbative aspects of gauge field theory and string theory. For
further work it may also be important to find the relation between the
duality symmetry for $Z_{M({\cal G)}}(k)$ given in (27) and $S$-duality.

Besides of the possible connection between M(atroid)-theory and M-theory
there is another interesting physical application of the present work. This
has to do with loop solutions of quantum gravity based in Ashtekar
formalism. It is known that the Witten's partition function provides a
solution of the Ashtekar constriants [47]. So, the duality symmetries (27)
also applies to such solutions. In other words, it seems that we have also
found a connection between matroid theory and loop solutions of quantum
canonical gravity.

\end{document}